\documentclass[twocolumn, unpublished,a4paper,accepted=2019-06-29]{quantumarticle}
\pdfoutput=1

\usepackage{amsmath,amsfonts,amssymb,bm,graphicx,verbatim,mathrsfs,units}
\usepackage[numbers,sort&compress]{natbib}

\usepackage{color,colordvi}
\definecolor{blue}{rgb}{0,0,1}
\definecolor{grey}{rgb}{0.6,0.6,0.6}

\usepackage{hyperref}

\begin{document}


\title{Fundamental limits on low-temperature quantum thermometry with finite resolution}

\author{Patrick P. Potts}
\email{patrick.potts@teorfys.lu.se}
\thanks{Formerly known as Patrick P. Hofer}
\orcid{0000-0001-6036-7291}
\author{Jonatan Bohr Brask}
\orcid{0000-0003-3859-0272}
\author{Nicolas Brunner}
\affiliation{Department of Applied Physics, University of Geneva, Switzerland}


\date{\today}

\begin{abstract}
While the ability to measure low temperatures accurately in quantum systems is important in a wide range of experiments, the possibilities and the fundamental limits of quantum thermometry are not yet fully understood theoretically. Here we develop a general approach to low-temperature quantum thermometry, taking into account restrictions arising not only from the sample but also from the measurement process. {We derive a fundamental bound on the minimal uncertainty for any temperature measurement that has a finite resolution. A similar bound can be obtained from the third law of thermodynamics. Moreover, we identify a mechanism enabling sub-exponential scaling, even in the regime of finite resolution. We illustrate this effect in the case of thermometry on a fermionic tight-binding chain with access to only two lattice sites, where we find a quadratic divergence of the uncertainty}. We also give illustrative examples of ideal quantum gases and a square-lattice Ising model, highlighting the role of phase transitions.


\end{abstract}



A sizable part of contemporary physics is focused on phenomena that occur at low temperatures \cite{altland:book}. While for some of these phenomena, finite temperatures constitute an undesirable nuisance (usually in the form of noise), in other scenarios temperature gradients are at the heart of the physics under investigation \cite{vinjanampathy:2016,goold:2016}. In both cases, accurately determining temperature is desirable for characterizing the examined processes and ultimately leads to an increase in control and a better understanding of the underlying physics \cite{carlospalacio:2016,pasquale:2018}.

If the precision is quantified by the relative error, it is not surprising that thermometry becomes more challenging as temperature is reduced. However, the state of the matter is even worse as it turns out that also the absolute error usually increases as temperature is reduced. Sensitively measuring low temperatures is therefore challenging. But how hard is it exactly? 

A number of previous studies considered specific physical models and found that it is exponentially hard, in the sense that the absolute error diverges exponentially at low temperatures \cite{stace:2010,brunelli:2011,brunelli:2012,sabin:2014,mehboudi:2015,guo:2015,hofer:2017prl,pasquale:2017,campbell:2017,campbell:2018,plodzien:2018,sone:2018}. In fact, this exponential scaling can be derived from very general arguments \cite{paris:2009,pasquale:2016,depalma:2017,correa:2015}, based on the energy spectra of the sample and the probe, and could therefore appear to represent a fundamental limit on thermometry. However, it was recently shown that {(using a measurement with infinite resolution)} a sub-exponential scaling is possible in a system of strongly coupled harmonic oscillators \cite{correa:2017}. This raises the question of what the fundamental limits to low-temperature quantum thermometry really are. In particular, when and how exponential scaling can be overcome. In addition to fundamental insight, addressing this question is also highly desirable given the importance of thermometry for quantum experiments \cite{giazotto:2006}.


With this goal in mind, we present a general approach to low-temperature quantum thermometry. {In addition to the limitations imposed by the system itself, we consider a constraint on the measurement, namely that measurements can only have finitely many outcomes (implying finite resolution). Under this constraint alone, we derive a fundamental bound on quantum thermometry.} Alternatively the same bound can in fact be recovered from the third law of thermodynamics, without making any assumptions on the measurements. More generally, our approach clearly identifies regimes featuring an exponentially diverging error, and those where sub-exponential scaling is possible even with finite measurement resolution. We show that the latter is possible in a physical system, namely a fermionic tight-binding chain with access to only two lattice sites. In this case, the uncertainty diverges only quadratically. Hence the same scaling as in the model of Ref.~\cite{correa:2017} is achieved, but here requiring only finite resolution. Furthermore, we apply our formalism to illustrative examples of ideal quantum gases and a square-lattice Ising model, highlighting the role of phase transitions.


\begin{figure*}[t]
	\centering
	\includegraphics[width=.9\textwidth]{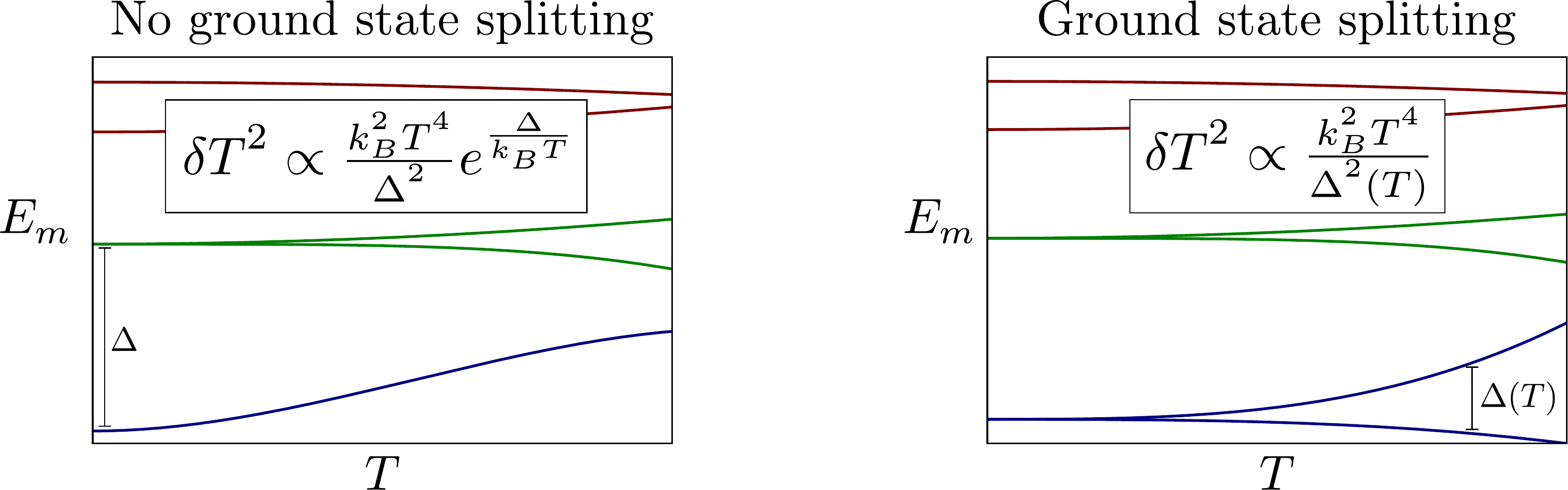}
	\caption{Illustration of different low-temperature scalings of the smallest possible error in thermometry. Given a measurement, an energy can be associated to each outcome, corresponding to the most likely system energy given that outcome. These energies define a temperature-dependent spectrum $E_m$ which in general differs from the spectrum of the system Hamiltonian. Left panel: as long as the ground-state energy does not exhibit a degeneracy lifting, the best achievable measurement error scales exponentially. Right panel: a degeneracy lifting of the ground-state as a function of temperature leads to polynomial scaling of the measurement error at low temperatures. Note that for the explicit scaling shown, we assumed that there is no term linear in $T$ in the lowest gap for both panels.}
	\label{fig:1}
\end{figure*}

\section{Context and main results}

Before we consider bounds that are imposed by limited experimental access, it is illustrative to consider the limitations imposed by the system itself. The resulting bounds are mostly based on the quantum Fisher information and provide a minimal variance for temperature estimation (under certain conditions that are specified below) \cite{paris:2009}. For thermal states in the canonical ensemble, such a bound is derived by Paris in Ref.~\cite{paris:2016} (see also Ref.~\cite{zanardi:2007}) and reads
\begin{equation}
\label{eq:boundparis}
\delta T^2\geq \frac{1}{\nu}\frac{k_BT^4}{\langle\hat{H}^2\rangle-\langle\hat{H}\rangle^2}=\frac{1}{\nu}\frac{k_BT^2}{C},
\end{equation}
where $\nu$ denotes the number of independent measurements involved in the estimation, $\hat{H}$ is the Hamiltonian, and $C$ the heat capacity of the system. This bound holds for all systems and it is interesting to note that a classical calculation by Landau and Lifshitz yields the exact same result \cite{landaulifshitz} (see also App.~\ref{app:classical}). Even though Eq.~\eqref{eq:boundparis} depends on the system of interest, we can use the third law of thermodynamics to obtain a bound that is system independent. To this end, we note that the heat capacity vanishes as $T\rightarrow 0$ (the relations between vanishing heat capacities and the Nernst postulate are discussed below and in  Ref.~\cite{callen:book}). This directly implies that the \textit{relative} error $\delta T^2/T^2$ necessarily diverges as $T\rightarrow 0$. We note that Eq.~\eqref{eq:boundparis} implies that temperature can be measured precisely whenever the heat capacity diverges. This can happen at phase transitions and is illustrated below.

For any system with a finite energy gap $\Delta$ between the ground and the first excited state, Eq.~\eqref{eq:boundparis} implies that the \textit{absolute} error in any temperature measurement diverges exponentially as absolute zero is approached {(i.e., $k_BT\ll\Delta$)} \cite{paris:2016}
\begin{equation}
\label{eq:boundparis2}
\delta T^2\geq \frac{1}{\nu}\frac{g_0}{g_1}\frac{k_BT^4}{\Delta^2}e^{\beta\Delta},
\end{equation}
where $\beta=1/(k_BT)$ denotes the inverse temperature and $g_0$ ($g_1$) the degeneracy of the ground (first excited) state. {The exponentially diverging error is a consequence of the fact that for temperatures far below the lowest gap, the system is essentially in the ground state, irrespective of the exact value of $T$.} Such an exponentially diverging error was found in various studies on low-temperature thermometry (see, e.g., \cite{stace:2010,brunelli:2011,brunelli:2012,sabin:2014,mehboudi:2015,guo:2015,correa:2015,pasquale:2016,depalma:2017,correa:2017,hofer:2017prl,campbell:2017,pasquale:2017,campbell:2018,plodzien:2018,sone:2018}). In particular, Eq.~\eqref{eq:boundparis2} is the relevant bound if a weakly coupled, small probe is used to determine the temperature of the system (assuming that the energy of the probe can be measured projectively). In that case, the probe thermalizes with the system and, due to the weak coupling, is described by a thermal state determined by the probe Hamiltonian. Note that the degeneracy factors in Eq.~\eqref{eq:boundparis2} imply that it is beneficial to use a probe with a highly degenerate excited state as discussed in Ref.~\cite{correa:2015}. 

The bound in Eq.~\eqref{eq:boundparis2} seems to imply that measuring cold temperatures is exponentially difficult in general. However, this is not necessarily the case. To see this, we discuss the limitations of Eq.~\eqref{eq:boundparis2}. The first limitation is the assumption of a finite gap between ground and excited state. While such a gap is strictly speaking always present as long as the system is of finite size, it can be far below any relevant energy scale (including the resolution of any energy measurement). For systems which can effectively be described by a continuous energy spectrum, the relevant bound thus cannot be given by Eq.~\eqref{eq:boundparis2}, {as the limit $k_BT\ll\Delta$ cannot be achieved}. This will be illustrated by several examples. In this case, one might be tempted to take a step back and resort to Eq.~\eqref{eq:boundparis}. For electronic systems for instance, the heat capacity usually vanishes linearly in $T$ \cite{ashcroft:book}. Equation \eqref{eq:boundparis} then implies that the error $\delta T^2$ also vanishes linearly in $T$. This brings us to the second limitation of Eqs.~\eqref{eq:boundparis} and \eqref{eq:boundparis2}: the assumption of complete accessibility. The bound in Eq.~\eqref{eq:boundparis}
can be saturated when performing projective measurements of the system energy (in the large $\nu$ limit) \cite{marzolino:2013,paris:2016}. This requires the possibility to experimentally distinguish between all (non-degenerate) eigenstates of the system Hamiltonian. For systems that are effectively described by a continuous spectrum, this is a hopeless task as it requires an infinite resolution in energy.

While Eqs.~\eqref{eq:boundparis} and \eqref{eq:boundparis2} give lower bounds on any obtained error, they might thus be of little practical use when the temperature estimation is bounded by the experimental access (see also Ref.~\cite{frowis:2016} for the influence of a limited detector resolution on the obtainable Fisher information). It is therefore highly desirable to have bounds on low-temperature thermometry that take into account limited access to the system. This includes the interesting scenario of determining temperature using a small probe that is coupled strongly to a large sample. In this case, the reduced state of the probe differs from a thermal state described by the probe Hamiltonian. As shown in Ref.~\cite{correa:2017}, the bound in Eq.~\eqref{eq:boundparis2} (with $\Delta$ being the gap of the probe), can then be surpassed. This is possible because the total system (sample and probe) is gapless. {For any total system that is \textit{not} gapless, an exponential divergence according to Eq.~\eqref{eq:boundparis2} cannot be avoided \cite{hovhannisyan:2018}.}

The main goal of this work is to provide a detailed investigation of low-temperature thermometry under limited access, resulting in a number of novel bounds on the associated measurement error. To this end, we consider measurements with finitely many outcomes. One can then assign an energy as well as a probability to each measurement outcome. These quantities replace the energies and occupation probabilities of the Hamiltonian in Eq.~\eqref{eq:boundparis} {and define a spectrum which determines the error in the temperature estimation. {The assumption of finite resolution implies that this spectrum cannot be treated as gapless.}
 If the spectrum can be approximated} as temperature-independent at low $T$, we immediately recover an exponentially diverging error analogous to Eq.~\eqref{eq:boundparis2}, see Fig.~\ref{fig:1} ${\rm (a)}$. However, the spectrum that is associated to the measurement is in principle temperature-dependent. This allows to overcome the exponential scaling of $\delta T^2$ by the following mechanism: Whenever the ground state exhibits a degeneracy at $T=0$ which is lifted at finite temperatures, the error $\delta T^2$ scales polynomially as $T\rightarrow 0$, see Fig.~\ref{fig:1} ${\rm (b)}$.
 {In this work, we show that such a sub-exponential scaling can be observed using measurements with finite resolution. This occurs in a tight-binding chain with measurement access restricted to only two lattice sites, leading to the scaling $\delta T^2\propto 1/T^2$. We note that the same scaling is also obtained in a system of strongly coupled harmonic oscillators using a measurement with infinite resolution \cite{correa:2017}.}
 
The tight-binding chain provides an upper bound on the best possible scaling that can be physically achieved. {Furthermore, we show that the restriction of finite resolution alone results in a fundamental lower bound on the precision of quantum thermometry. Interestingly, this bound turns out to be equivalent to the one implied by the third law of thermodynamics, and is less restrictive than the scaling we obtain for the tight-binding chain. While the relative error of any temperature measurement necessarily diverges, our bound in principle allows for the absolute error to vanish in the limit $T\rightarrow 0$. It is thus an open question whether there exists a physical system where temperature can be determined perfectly using measurements with finite resolution.} {We note that while in principle any measurement has a finite resolution, there may be scenarios where an infinite resolution can safely be assumed, i.e., where its associated spectrum is approximately gapless.}

The rest of this paper is structured as follows: in Sec.~\ref{sec:sett}, we introduce the relevant quantities and the applied formalism. The low-temperature bounds are then derived in Sec.~\ref{sec:bounds}, identifying the condition under which sub-exponential scaling can be achieved. In Sec.~\ref{sec:examples}, we illustrate our results considering ideal quantum gases. In Sec.~\ref{sec:examplesA}, we consider finite systems and show how they behave as infinite systems at elevated temperatures, illustrating the inadequacy of Eq.~\eqref{eq:boundparis2} for systems with small gaps. In Sec.~\ref{sec:examplesB}, we discuss a tight-binding chain with restricted access, where polynomial scaling is achieved. We illustrate the effect of phase transitions on low-temperature thermometry in finite systems in Sec.~\ref{sec:phaset}. We conclude in Sec.~\ref{sec:conclusions} discussing a number of open questions. 
 
\section{Framework for temperature measurements}
\label{sec:sett}
Here we consider the scenario where temperature is determined through a general quantum measurement described by a positive-operator valued measure (POVM) with $M$ elements $\hat{\Pi}_m$. We note that POVMs provide the most general measurements in quantum theory \cite{nielsen:book}. {They include scenarios where auxiliary systems interact in an arbitrary way with the system of interest} as well as measurements that exploit quantum coherence \cite{stace:2010,monras:2011,martin:2013,sabin:2014,jevtic:2015,mehboudi:2015,johnson:2016,mancino:2017,razavian:2018,cavina:2018}.
Each of the POVM elements corresponds to an outcome $m$ which will be observed with probability
\begin{equation}
\label{eq:pm}
p_m={\rm Tr}\left\{\hat{\Pi}_m\hat{\rho}_T\right\}=\sum_n \frac{e^{-\beta\mathcal{E}_n}}{Z}\langle n|\hat{\Pi}_m|n\rangle,
\end{equation}
where $\hat{\rho}_T$ denotes the thermal state $\hat{\rho}_T={e^{-\beta\hat{H}}}/{Z}$ with $Z={\rm Tr}\{e^{-\beta\hat{H}}\}$
and $\hat{H}|n\rangle=\mathcal{E}_n|n\rangle$.
For simplicity, we consider the canonical ensemble. Extensions to the grand-canonical ensemble, including a temperature-dependent chemical potential, are discussed in App.~\ref{app:mu}. {Throughout, we consider the POVM elements to be temperature-independent.}

The amount of information on temperature that is encoded in the probability distribution $p_m$ is given by the Fisher information \cite{paris:2009}
\begin{equation}
\label{eq:fisher}
F_T=\sum_{m=0}^{M-1}\frac{\left(\partial_T p_m\right)^2}{p_m}.
\end{equation}
We note that the Fisher information explicitly depends on the POVM. Through the Cram\'er-Rao bound, the Fisher information gives a lower bound on the variance of any unbiased estimator for $T$ \cite{cramer:book}
\begin{equation}
\label{eq:cramerrao}
\delta T^2\geq\frac{1}{\nu F_T},
\end{equation}
where $\nu$ denotes the number of independent measurement rounds. This variance will serve as our quantifier for the uncertainty of the temperature measurement.

We now define an energy for each measurement outcome 
\begin{equation}
\label{eq:em}
E_m=\frac{1}{p_m}{\rm Tr}\left\{\hat{\Pi}_m\hat{H}\hat{\rho}_T\right\}=
\frac{\sum_n \mathcal{E}_ne^{-\beta\mathcal{E}_n}\langle n|\hat{\Pi}_m|n\rangle}{\sum_{n'} e^{-\beta\mathcal{E}_{n'}}\langle n'|\hat{\Pi}_m|n'\rangle},
\end{equation}
which can be interpreted as the optimal guess of the system energy (before the measurement) given the outcome $m$. Note that if the POVM elements are projectors onto the eigenstates of the Hamiltonian, then the energies defined in Eq.~\eqref{eq:em} coincide with the eigenenergies of the system, i.e. $E_m=\mathcal{E}_m$. The Fisher information can be written as the variance of the spectrum defined by Eqs.~\eqref{eq:pm} and \eqref{eq:em}
\begin{equation}
\label{eq:fishercl}
F_T=\frac{1}{k_B^2T^4}\left[\sum_{m=0}^{M-1}p_mE^2_m-\left(\sum_{m=0}^{M-1}p_mE_m\right)^2\right],
\end{equation}
where we used Eq.~\eqref{eq:fisher} with
\begin{equation}
\label{eq:partpm}
\partial_T p_m =\left({E}_m-\sum_{n=0}^{M-1}p_n {E}_n\right)\frac{p_m}{k_BT^2}.
\end{equation}
As we discuss in detail below, defining the spectrum associated to a given POVM allows us to group the POVMs into different classes. These classes are defined through the thermal behavior of the low-energy spectrum and they lead to different bounds on the Fisher information. {We stress that the energies $E_m$ are only a convenient tool. Their definition does in no way influence the generality of the results.}

A measurement independent quantity, the quantum Fisher information (QFI), can be obtained by maximizing the Fisher information over all possible POVMs. In the case of determining temperature (in the canonical ensemble), the optimal POVM corresponds to a projective measurement of the energy of the system and the QFI is determined by the variance of the system energy \cite{paris:2016}
\begin{equation}
\label{eq:qfi}
\begin{aligned}
\mathcal{F}_T&=\frac{\langle \hat{H}^2\rangle -\langle \hat{H}\rangle^2}{k_B^2T^4}=\frac{\partial_\beta^2\ln Z}{k_B^2T^4}\\& = \frac{\partial_T S}{k_BT} = \frac{\partial_T\langle \hat{H}\rangle}{k_B T^2},
\end{aligned}
\end{equation}
where we have given the QFI in terms of different thermodynamic quantities for future reference and we introduced the von Neumann entropy $S=-k_B{\rm Tr}\{\hat{\rho}_{T}\ln \hat{\rho}_{T}\}$. We note that for thermal states, the von Neumann entropy unambiguously corresponds to the thermodynamic entropy \cite{jaynes:1957,vinjanampathy:2016}.

Using the definition of the heat capacity
\begin{equation}
\label{eq:heatcap}
C=\frac{\delta Q}{dT} =T\partial_T S,
\end{equation}
where $\delta Q$ is the infinitesimal amount of heat required for a temperature change of $dT$, we can write the QFI as
\begin{equation}
\label{eq:qfic}
\mathcal{F}_T=\frac{C}{k_BT^2}.
\end{equation}
We note that the heat capacity is usually defined at constant volume or pressure. The above relations hold no matter which quantity is being held constant as temperature is varied (i.e., if volume is held constant, $C$ denotes the heat capacity at constant volume and similarly for any other choice). The last expression implies that temperature can be measured precisely when the heat capacity diverges. This can happen at phase transitions, where the heat capacity exhibits a singular behavior. Using phase transitions for thermometry has a long history \cite{giazotto:2006,zanardi:2008,carlospalacio:2016} (think of the historical definition of the Celsius scale using the freezing and boiling points of water) and is illustrated below with the examples of Bose-Einstein condensation (where $C$ exhibits a cusp) and the two-dimensional Ising model (where $C$ diverges at the critical temperature).

From Eq.~\eqref{eq:heatcap}, one infers that the heat capacity vanishes as $T\rightarrow0$ as long as the derivative $\partial_T S$ remains finite. While the Nernst postulate only enforces the entropy to be constant at $T=0$ (implying $\partial_X S|_{T=0}=0$) the assumption of a finite derivative $\partial_T S$ is often included in the third law of thermodynamics \cite{callen:book}. This implies the following bound on the QFI
\begin{equation}
\label{eq:bound3rdlaw}
\lim_{T\rightarrow0}T^2\mathcal{F}_T\rightarrow0.
\end{equation}
It follows that the relative error $\delta T^2/T^2$ in any temperature measurement has to diverge as absolute zero is approached [cf.~Eq.~\eqref{eq:cramerrao}]. In addition to the unattainability principle of reaching absolute zero \cite{masanes:2017,wilming:2017}, the third law thus implies an unattainability principle for precisely measuring temperatures close to absolute zero. {We note that idealized scenarios (e.g. a free particle not constrained by any container) may result in finite heat capacities at $T=0$.} Equation \eqref{eq:bound3rdlaw} constitutes our first and weakest bound on low temperature thermometry. In the next section, we will give stronger bounds which are based on restrictions on accessibility.

We note that all considerations that are measurement independent, i.e. all above considerations based on the QFI, can be reproduced by a classical calculation following Landau and Lifshitz \cite{landaulifshitz}. For completeness, this derivation is reproduced in App.~\ref{app:classical}.

\section{Bounds on low temperature thermometry}
\label{sec:bounds}
{To investigate bounds on low-temperature thermometry, we consider an arbitrary POVM} with finitely many outcomes. {This restriction is discussed in more detail at the end of this section} and it implies that the spectrum $E_m$ is discrete. To find bounds on the Fisher information, we write the formal solution to Eq.~\eqref{eq:partpm} as
\begin{equation}
\label{eq:pmformal}
p_m = \frac{e^{\int dT \Delta_m(T)/(k_BT^2)}}{\sum_{n=0}^{M-1}e^{\int dT \Delta_n(T)/(k_BT^2)}},
\end{equation}
where $\Delta_m=E_m-E_0\geq 0$.
We now show how the low-temperature behavior of the spectrum $E_m$ leads to different bounds on the error in temperature estimation.

\subsection{No degeneracy splittings}

Let us first consider the case where the spectrum $E_m$ can be approximated as temperature independent at low $T$. In this case, the probabilities reduce to the Boltzmann form
\begin{equation}
\label{eq:pmboltzmann}
p_m=\frac{g_me^{-\beta \Delta_m}}{\sum_{n}g_ne^{-\beta \Delta_n}},
\end{equation}
where the factors $g_m$ are the temperature-independent integration constants in Eq.~\eqref{eq:pmformal}. They are determined by Eq.~\eqref{eq:pm}. Without loss of generality, we can absorb all degeneracies in the factors $g_m$. Relabeling the energies then results in a non-degenerate spectrum where $\Delta_j>\Delta_i$ for $j>i$. Inserting Eq.~\eqref{eq:pmboltzmann} in Eq.~\eqref{eq:fishercl} results in the low temperature behavior
\begin{equation}
\label{eq:fisherexp}
F_T=\frac{g_1}{g_0}\frac{\Delta_1^2}{k_B^2T^4}e^{-\beta \Delta_1}+\mathcal{O}(e^{-\beta{\rm min}(2\Delta_1,\Delta_2)}/T^4).
\end{equation}
The temperature measurement is then necessarily accompanied by an exponentially divergent error as $T\rightarrow0$ in complete analogy to Eq.~\eqref{eq:boundparis2} with the only difference that the gap $\Delta_1$ might be unrelated to and much bigger than the lowest gap in the spectrum of the system Hamiltonian. In case the implementable POVMs are far from the optimal one, the temperature measurement might perform considerably worse than predicted by Eqs.~\eqref{eq:boundparis} and \eqref{eq:boundparis2}. {This is of particular relevance when the optimal POVM requires an infinite resolution.}

Next, we relax the assumption of temperature independent energies and we assume that the $E_m$ can be Taylor expanded around $T=0$
\begin{equation}
\label{eq:entaylor}
\Delta_m=\Delta_{m,0}+k_B\sum_{l=1}^{\infty}\Delta_{m,l}T^l.
\end{equation}
In this case, the probabilities read
\begin{equation}
\label{eq:probagen}
p_m=\frac{g_mT^{\Delta_{m,1}}e^{-\beta\Delta_{m,0}}e^{\sum_{l=1}^\infty\Delta_{m,l+1}T^l/l}}{\sum_n g_nT^{\Delta_{n,1}}e^{-\beta\Delta_{n,0}}e^{\sum_{l=1}^\infty\Delta_{n,l+1}T^l/l}},
\end{equation}
where $g_mT^{\Delta_{m,1}}$ is a dimensionless quantity. 
{We note that by virtue of Weierstrass' approximation theorem \cite{weierstrass:1885}, any continuous $\Delta_m(T)$ can be approximated to arbitrary precision by Eq.~\eqref{eq:entaylor}. To investigate the scaling of the Fisher information, we implicitly assumed that the Fisher information is a continuous function of $T$, implying that also $\Delta_m$ are continuous functions, justifying the use of Eq.~\eqref{eq:entaylor}.}

We now consider the case where no degeneracies are lifted as a function of temperature [i.e., $\Delta_i(0)=\Delta_j(0) \Rightarrow \Delta_i(T)=\Delta_j(T)$]. In this case, we can again absorb all degeneracies in the factors $g_m$ and consider a gapped, non-degenerate spectrum where $\Delta_j>\Delta_i$ for $j>i$. We note that we only consider temperatures that are low enough such that no level crossings have to be taken into account. Inserting Eqs.~\eqref{eq:entaylor} and \eqref{eq:probagen} in Eq.~\eqref{eq:fisher} and only keeping the largest term in the limit $T\rightarrow0$ then results in
\begin{equation}
\label{eq:fisherexp2}
F_T=\frac{g_1}{g_0}\frac{\Delta_{1,0}^2}{k_B^2T^{4-\Delta_{1,1}}}e^{-\beta\Delta_{1,0}}+\mathcal{O}(e^{-\beta\Delta_{1,0}}/T^{3-\Delta_{1,1}}).
\end{equation}
We thus see that the temperature dependence of $\Delta_1$ can slightly modify the low-temperature scaling of the Fisher information. However, we still find an exponentially diverging error. 

\subsection{Degeneracy splittings}
{We now turn to the case where a sub-exponential scaling can be observed.} To this end, we consider the case of a ground-state which exhibits a degeneracy splitting as a function of temperature. At low temperatures, it is then sufficient to consider the ground-state manifold. All additional terms of the Fisher information vanish exponentially as $T\rightarrow 0$. We thus have $\Delta_{m,0}=0$ for all considered energies. As temperature increases, the ground-state will split into different levels. Let the largest gap that opens up grow with $T^l$. All gaps are then approximated as
\begin{equation}
\label{eq:smalldel}
\Delta_j=k_B\Delta_{j,l}  T^l +\mathcal{O}(T^{l+1}),
\end{equation}
i.e., gaps with $\Delta_{j,l}=0$ are neglected and the corresponding energies are absorbed into $g_0$. Let us consider the case $l>1$. From Eqs.~\eqref{eq:probagen}, \eqref{eq:smalldel}, and \eqref{eq:fishercl}, we find a polynomial low temperature behavior
\begin{widetext}
\begin{equation}
\label{eq:fisherpol}
F_T=T^{2(l-2)}\left[\sum_m \Delta_{m,l}^2\frac{g_m}{\sum_m g_m}-\left(\sum_{m}\Delta_{m,l}\frac{g_m}{\sum_m g_m} \right)^2\right]+\mathcal{O}(T^{3l-5}).
\end{equation}
\end{widetext}
{In contrast to the above discussed gapless systems, where the QFI corresponds to a POVM with an infinite resolution, the sub-exponential nature of Eq.~\eqref{eq:fisherpol} is accessible under the restriction of finite resolution.}
From the last equation, we find that for $l=2$, the Fisher information remains constant at $T=0$ implying that the absolute error of a low temperature measurement does not necessarily diverge. For all higher $l$, we find a polynomial divergence of the measurement error. Finally, let us consider the case $l=1$ for which we find
\begin{equation}
\label{eq:fisherlin}
F_T=\frac{g_1}{g_0}\Delta_{1,1}^2T^{\Delta_{1,1}-2}+\mathcal{O}(T^{{\rm min}(2\Delta_{1,1},\Delta_{2,1})-2}).
\end{equation}
As $\Delta_{1,1}$ tends to zero, we find that we can obtain any scaling which respects the third law of thermodynamics, i.e. which fulfills Eq.~\eqref{eq:bound3rdlaw}. {It is interesting to note that imposing finite resolution results in the exact same ultimate scaling that can not be overcome as the third law of thermodynamics. Equation \eqref{eq:fisherlin} thus provides a rigorous lower bound on the best scaling that can be observed with measurements of finite resolution.}

Finally, we consider the case where only the excited states exhibit a degeneracy splitting as a function of temperature. At low temperatures, it is sufficient to consider the first excited state. If the first excited state is $p$-fold degenerate at $T=0$, we have $\Delta_{j,0}=\Delta_{1,0}$ for all $0<j\leq p$. However, $\Delta_{j,1}$ can be different for each $j$. A similar derivation as for Eq.~\eqref{eq:fisherexp2} shows that in this case, the largest term in the limit $T\rightarrow 0$ is still given by Eq.~\eqref{eq:fisherexp2}. However, the terms that are dropped are at least of the order of $\mathcal{O}(e^{-\beta\Delta_{1,0}}/T^{4-\Delta_{2,1}})$.

To summarize, we consider POVMs with finitely many outcomes. In this case, the only way to overcome an exponentially diverging error in low-temperature thermometry is to use a POVM such that the ground state splits into multiple levels at finite temperatures [as illustrated in Fig.~\ref{fig:1} ${\rm (b)}$]. The thermal variation of the occupation numbers of the ground-state manifold are then not exponentially small which allows for more precise thermometry. This insight allows for designing experimentally accessible POVMs for precise low-temperature thermometry.
Below we provide a physical example for a system where the Fisher information scales as $T^2$ [i.e., $l=3$ in Eq.~\eqref{eq:fisherpol}]. {This example shows that sub-exponential scaling can be obtained under the restriction of finite resolution. We note that the same scaling was found in Ref.~\cite{correa:2017} for a different system and a measurement with infinitely many outcomes.} It remains an open question if a better scaling can be achieved in a physical system. 

\subsection{Relevant low temperature limit}
{Finite sized systems always exhibit a finite gap between the ground and the first excited states. Therefore, Eq.~\eqref{eq:boundparis2} always becomes relevant at sufficiently low temperatures. However, in many systems of experimental interest, the gaps between the low-lying states are by far the smallest energy scales of the problem and we can predict all experimental observables by formally letting the gaps go to zero. In this case, any scaling allowed by Eq.~\eqref{eq:bound3rdlaw} can in principle be achieved.
	
{Just as a large but finite system may be assumed to have no gap in the spectrum of the Hamiltonian, there may be POVMs which have corresponding spectra that, for all practical purposes, can be treated as gapless. This corresponds to scenarios where temperature is always sufficiently big such that the terms that are dropped in the bounds discussed in this section remain relevant.} The relevant low-temperature limit for the provided scalings is thus
	\begin{equation}
	\label{eq:lowtrel}
	\Delta\ll k_BT\ll \omega,\,\Delta_1.
	\end{equation}
	Here $\Delta$ denotes the lowest gap in the spectrum of the Hamiltonian, $\omega$ stands for any energy scale in the system that does not vanish as $\Delta\rightarrow 0$, and $\Delta_1$ corresponds to the lowest gap in the spectrum of the considered POVM.}

\begin{figure*}
	\centering
	\includegraphics[width=\textwidth]{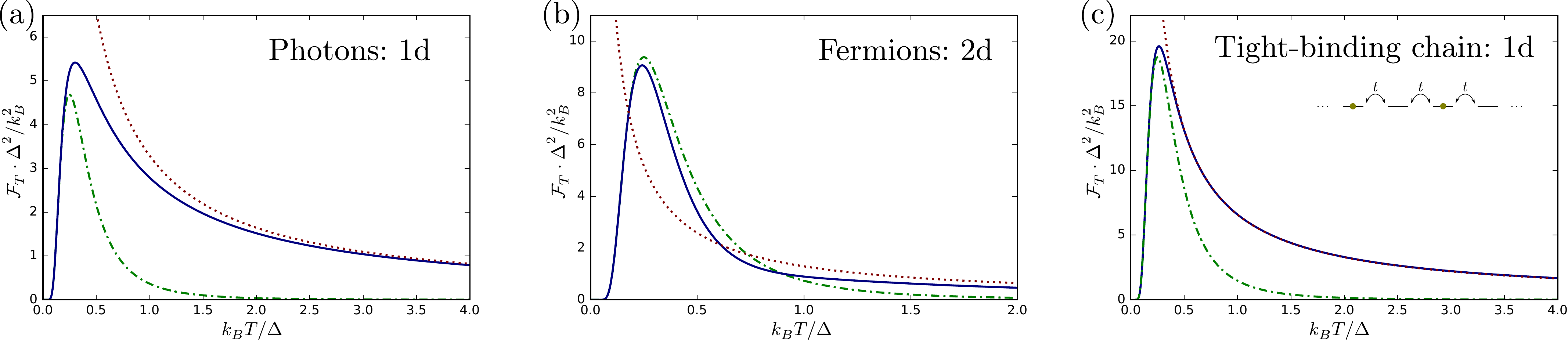}
	\caption{Quantum Fisher information for different systems. In all panels blue (solid) is the exact expression, green (dash-dotted) the exponential low-temperature limit given [cf.~Eq.~\eqref{eq:qfiltfb}], and red (dotted) the continuum approximation valid in the thermodynamic limit (where $L\rightarrow\infty$ and $\Delta\rightarrow0$). ${\rm (a)}$ One dimensional gas of photons contained in a space of length $L$. The smallest gap is $\Delta=\pi/L$ and the first excited state is non-degenerate. ${\rm (b)}$ Two dimensional gas of non-interacting fermions of mass $m$ contained in an area $L^2$. Here the smallest gap is given by $\Delta=\pi^2/(4mL^2)$ and the first excited state is two-fold degenerate. ${\rm (c)}$ Fermionic, one-dimensional tight-binding chain with hopping strength $t$, periodic boundary conditions, and $N=500$ sites. The smallest gap in this system is given by $\Delta=4\pi t/N$ and the first excited state is four fold degenerate (addition/removal of a right-/left-mover). In all three panels, the QFI diverges as $1/T$ in the continuum approximation. Overall, these results show that for temperatures above $\Delta$ the system can be effectively described as being gapless. For sufficiently small $\Delta$, the exponential low-temperature limit becomes thus irrelevant for all attainable temperatures.}
	\label{fig:examples_gapped}
\end{figure*}

\section{Examples}
\label{sec:examples}
In this section, we illustrate our results in a number of systems. Before turning to a system which exhibits a sub-exponential scaling of the Fisher information due to ground-state splitting, we investigate the quantum Fisher information in finite size systems. {In all examples, we consider a system described by a thermal state (in the canonical or grand-canonical ensemble) and we aim to determine the corresponding temperature.}

\subsection{Finite vs. infinite systems}
\label{sec:examplesA}
Here we illustrate the bounds given in Eqs.~\eqref{eq:boundparis} and \eqref{eq:boundparis2} for finite systems. At sufficiently low temperatures, we find the universality expressed by Eq.~\eqref{eq:boundparis2}, which holds for any gapped system. However,
for sufficiently small gaps (large systems), the discrete nature of the spectrum can be neglected and the bound in Eq.~\eqref{eq:boundparis2} becomes irrelevant for all attainable temperatures. 


We focus on non-interacting particles described by a spectrum of single-particle energies $\varepsilon_k$ and a chemical potential $\mu$ (for a detailed discussion on the quantum Fisher matrix with respect to $\beta$ and $\mu$, see Ref.~\cite{marzolino:2013}). 
In this case, the partition function in the grand-canonical ensemble reads for fermions
\begin{equation}
\label{eq:zmuf}
\ln Z_\mu^F=\sum_k \ln\left[1+e^{-\beta(\varepsilon_k-\mu)}\right],
\end{equation}
and for bosons
\begin{equation}
\label{eq:zmub}
\ln Z_\mu^B=-\sum_k \ln\left[1-e^{-\beta(\varepsilon_k-\mu)}\right].
\end{equation}
From Eq.~\eqref{eq:qfimut}, we find for fermions
\begin{equation}
\label{eq:qfif}
\mathcal{F}_T^F = \frac{1}{4k_B^2T^4}\sum_k\frac{[\varepsilon_k-\mu+T\partial_T\mu]^2}{\cosh^2\left[(\varepsilon_k-\mu)/(2k_BT)\right]},
\end{equation}
and bosons
\begin{equation}
\label{eq:qfib}
\mathcal{F}_T^B = \frac{1}{4k_B^2T^4}\sum_k\frac{[\varepsilon_k-\mu+T\partial_T\mu]^2}{\sinh^2\left[(\varepsilon_k-\mu)/(2k_BT)\right]}.
\end{equation}
In the low temperature limit, the last expressions both reduce to (assuming that $\partial_T\mu$ remains finite as $T\rightarrow0$)
\begin{equation}
\label{eq:qfiltfb}
\mathcal{F}_{T\rightarrow0}=\frac{g\Delta^2}{k_B^2T^4}e^{-\beta\Delta},
\end{equation}
in agreement with Eq.~\eqref{eq:boundparis2}. Here $\Delta={\rm min}_k|\varepsilon_k-\mu|$ denotes the smallest gap and $g$ its degeneracy. Note that for bosons $\varepsilon_k-\mu>0$, which implies that the ground-state is always given by zero bosons (i.e. non-degenerate) while the first excited state contains one boson. For fermions, the first excited state can either contain one extra fermion or one extra hole (missing fermion with $\varepsilon_k-\mu<0$). Note that the presence of single-particle energies equal to the chemical potential imply a degenerate ground-state. However, the same degeneracy is present in all excited states (we can always remove or add particles with $\varepsilon_k-\mu=0$ without changing the energy). This degeneracy therefore cancels in the QFI which is the reason that there is no ground-state degeneracy in Eq.~\eqref{eq:qfiltfb}.

\subsubsection{Photons}
As a first example, we consider (polarized) photons in a container with hard walls and a cubic volume $L^d$, where $d$ denotes the spatial dimension. The single-particle energies are then given by $\varepsilon_k=ck$ (we set $\hbar=1$ throughout the paper),
where $c$ denotes the speed of light and the wave number $k$ can take on the values
\begin{equation}
\label{eq:kphot}
k=\frac{\pi}{L}\sqrt{\sum_{i=1}^d n_i^2},
\end{equation}
with $n_i$ being positive integers. From Eq.~\eqref{eq:qfib} (with $\mu=0$), we find that the low temperature behavior of the QFI is given by Eq.~\eqref{eq:qfiltfb} with $g=1$ and $\Delta=c\sqrt{d}\pi/L$. 

For a macroscopic system, i.e., in the thermodynamic limit where $L\rightarrow\infty$, the gap $\Delta$ becomes vanishingly small and we can replace the sum in Eq.~\eqref{eq:qfib} with an integral. Doing so results in the QFI
\begin{equation}
\label{eq:qficontph}
\begin{aligned}
\mathcal{F}_T&=\frac{L^d}{4k_B^2T^2}\int\frac{d^dk}{(2\pi)^d}\frac{c^2k^2}{\sinh^2[ck/(2k_BT)]}\\&=\eta_d(k_BL/c)^dT^{d-2},
\end{aligned}
\end{equation}
where $\eta_1=\pi/3$, $\eta_2=3\zeta(3)/\pi$, and $\eta_3=2\pi^2/15$ with $\zeta(x)$ denoting the Riemann-zeta function. As expected, the QFI scales with the system size $L^d$ \cite{marzolino:2013}. The QFI for photons in one dimension is plotted in Fig.~\ref{fig:examples_gapped} ${\rm (a)}$, together with its low-temperature behavior and the thermodynamic limit. We find that for temperatures as low as $k_BT\gtrsim 2\Delta$, the QFI is well captured by the expression for the thermodynamic limit given in Eq.~\eqref{eq:qficontph}.
This expression diverges for small temperatures, allowing for precise thermometry in principle. However, this implies having access to an infinite system and being able to measure its energy with infinite precision (i.e., a POVM with infinite outcomes). In practice, the precision of any temperature measurement will in this case be limited by the classical Fisher information of the best experimentally implementable POVM.

\subsubsection{Massive particles}
As a next example, we consider a gas of massive, non-interacting particles in a hard wall container. The single-particle energies then read $\varepsilon_k={k^2}/{2m}$, where the quantum number $k$ can take on the values given in Eq.~\eqref{eq:kphot}. The QFI is then given by Eqs.~\eqref{eq:qfib} and \eqref{eq:qfif} for bosons and fermions respectively. The low temperature limit is again of the form given in Eq.~\eqref{eq:qfiltfb}, where the gap $\Delta$ and the degeneracy $g$ depend on the chemical potential and on the statistics of the particles (bosonic or fermionic).

The thermodynamic limit of massive particles is discussed in detail in App.~\ref{app:tdlimit}. For fermions with a positive and fixed chemical potential, we find that the QFI in the thermodynamic limit scales as $1/T$ at low temperatures, implying a heat capacity which vanishes linearly in $T$. This is illustrated in Fig.~\ref{fig:examples_gapped} ${\rm (b)}$, where we show the QFI for non-interacting fermions in two-dimensions with a fixed chemical potential. As for photons in one dimension, we find a diverging QFI in the limit $\Delta\rightarrow 0$ which provides a good description of the system for temperatures $k_BT\gtrsim 2\Delta$.

\begin{figure}[h!]
	\centering
	\includegraphics[width=.9\columnwidth]{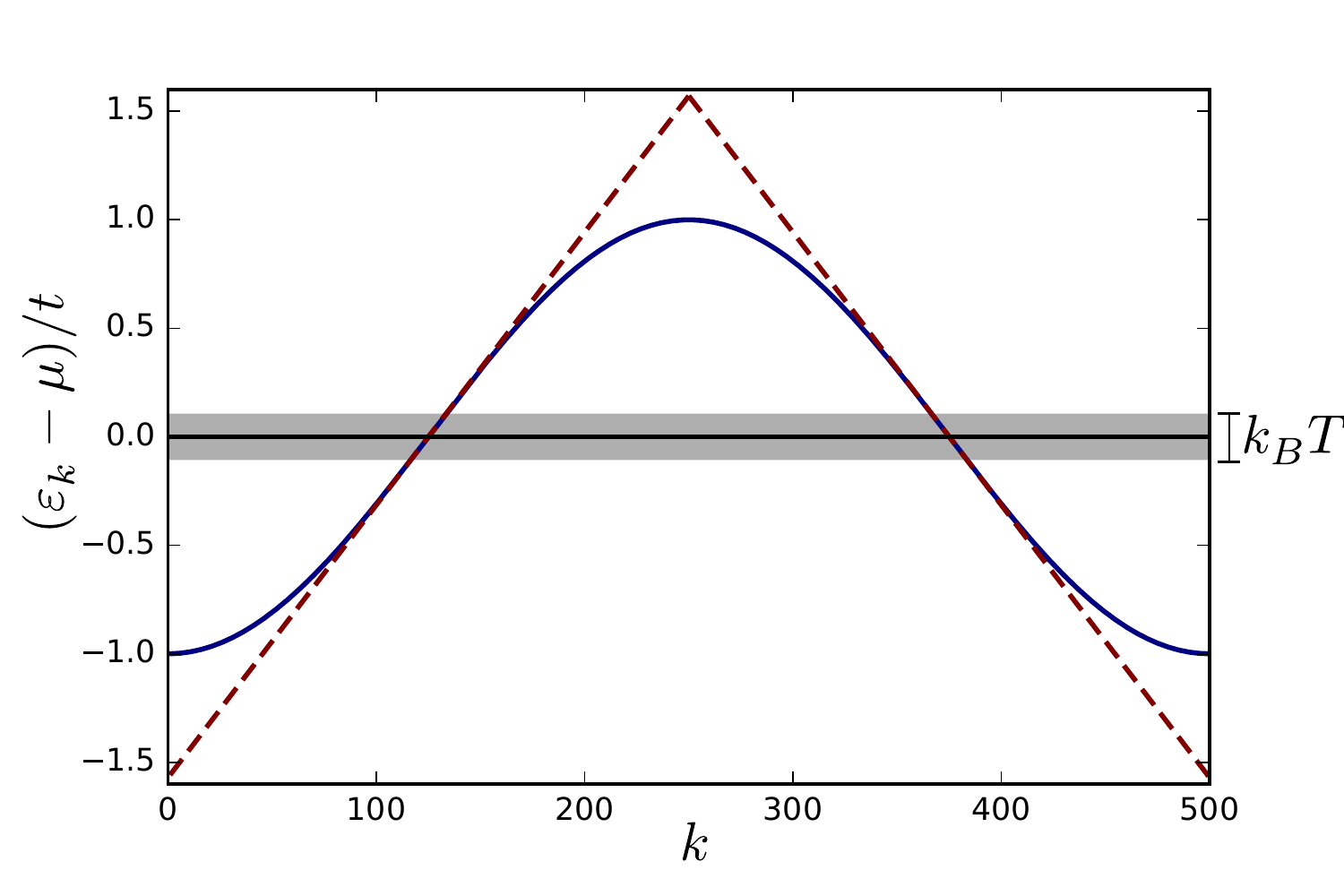}
	\caption{Spectrum of fermionic tight-binding chain with linear approximation for $\mu=\varepsilon$. Blue (solid): Correct spectrum as a function of $k$. Red (dashed): Linear approximation. For physical observables, only the partially filled state around the chemical potential are relevant. This is illustrated by the grey shading for $k_BT=0.2t$. Here, $N=500$.}
	\label{fig:spectrum}
\end{figure}

\subsubsection{Tight-binding chain}
As a next example, we consider a fermionic, one-dimensional tight binding chain described by the Hamiltonian
\begin{equation}
\label{eq:hamtb}
\hat{H}=\sum_{j=1}^N\varepsilon\hat{c}^\dagger_j\hat{c}_j - t\sum_{j=1}^{N}\left(\hat{c}_{j+1}^\dagger\hat{c}_j+\hat{c}_{j}^\dagger\hat{c}_{j+1}\right),
\end{equation}
where $\hat{c}_j$ annihilates a fermion on site $j$, $\varepsilon$ denotes the on-site energy (which is the same for all sites), $t$ the hopping strength, and $N$ the number of sites. We consider periodic boundary conditions ($\hat{c}_{j+N}=\hat{c}_j$), i.e., we consider a tight-binding chain that is arranged into a ring. Below, we will be interested in the thermodynamic limit $N\rightarrow\infty$, where the boundary conditions have a negligible effect. Diagonalizing Eq.~\eqref{eq:hamtb} results in
\begin{equation}
\label{eq:hamdiagtb}
\hat{H}=\sum_{k=1}^N\varepsilon_k\hat{c}^\dagger_k\hat{c}_k,
\end{equation}
with the eigenenergies $\varepsilon_k=\varepsilon-2t\cos( k {2\pi}/{N})$
and the eigenmodes
\begin{equation}
\label{eq:eigenstatestb}
\hat{c}_k =\frac{1}{\sqrt{N}}\sum_{j=1}^Ne^{ijk\frac{2\pi}{N}}\hat{c}_j.
\end{equation}
The QFI can then be evaluated using Eq.~\eqref{eq:qfif}. For a fixed chemical potential, the QFI is plotted in Fig.~\ref{fig:examples_gapped} ${\rm (c)}$.

In the following, we consider a particularly simple scenario which allows us to make some analytical progress. We fix the chemical potential at $\mu=\varepsilon$ (half-filling) and we focus on the regime $k_BT\ll t$. In this case, all (low-frequency) observables are determined by the partially filled states which are within $k_BT$ around the chemical potential. This allows us to linearize the single-particle energies $\varepsilon_k-\mu\simeq 2t\kappa$ with
\begin{equation}
\label{eq:kappa}
\kappa=\begin{cases}
k\frac{2\pi}{N}-\frac{\pi}{2},\hspace{.5cm}{\rm for}\,\, 1\leq k\leq\frac{N}{2}\\
\frac{3\pi}{2}-k\frac{2\pi}{N},\hspace{.4cm}{\rm for}\,\, \frac{N}{2}< k\leq N.
\end{cases}
\end{equation}
The last approximation is illustrated in Fig.~\ref{fig:spectrum}. 

With the help of this approximation, we can write the QFI in the thermodynamic limit ($N\rightarrow\infty$) as
\begin{equation}
\mathcal{F}_T=\frac{t^2N}{\pi k_B^2T^4}\int_{-\infty}^{\infty}d\kappa \frac{\kappa^2}{\cosh^2(\kappa t/k_BT)}=\frac{\pi k_B N}{6t T},
\end{equation}
where we extended the integral from the interval $[-\pi/2,\pi/2]$ to $[-\infty,\infty]$. Again, we find that the QFI is linear in the system size ($N$) and diverges as $1/T$ (just like a gas of massive fermions with a chemical potential within the spectrum). The QFI in the thermodynamic limit is plotted in Fig.~\ref{fig:examples_gapped} ${\rm (c)}$. As for the above examples, the QFI in the thermodynamic limit is only relevant if energy measurements with arbitrary precision are available.

\begin{figure*}[t]
\centering
\includegraphics[width=.9\textwidth]{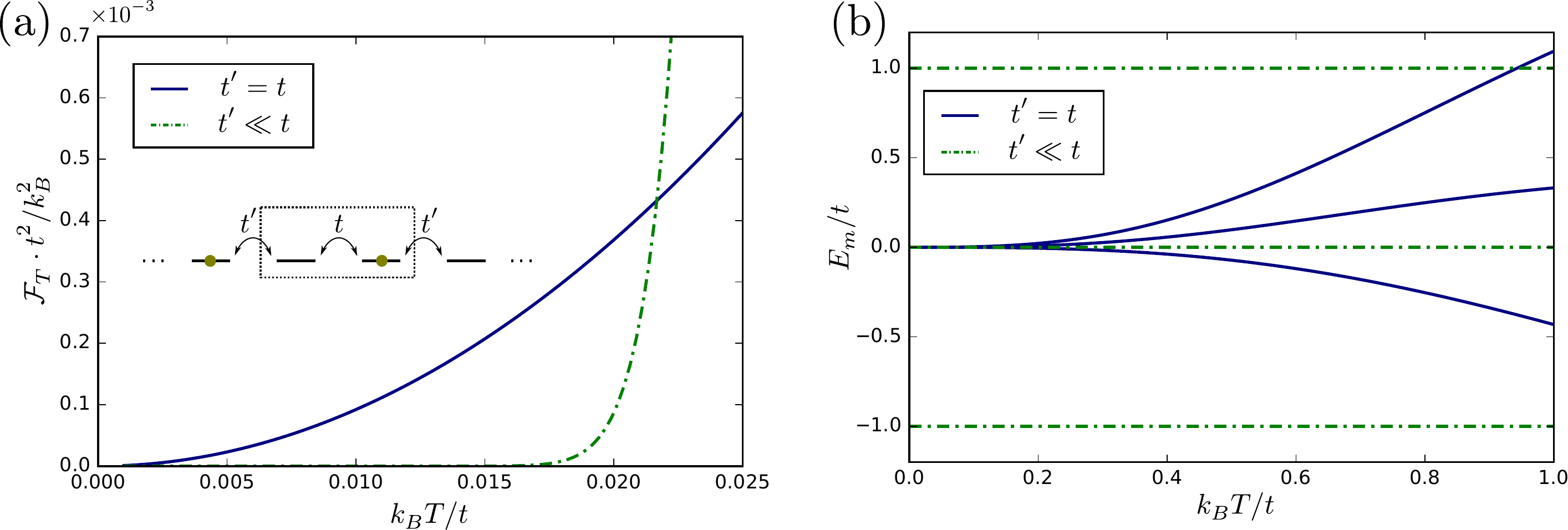}
\caption{Comparison between reduced states describing two sites of an infinite, fermionic tight-binding chain. ${\rm (a)}$ Quantum Fisher information. Green (dash-dotted): for a weak coupling between the two sites and the rest of the chain, we find the usual exponential scaling at low temperatures [cf.~Eq.~\eqref{eq:qfitb22}]. Blue (solid): if the coupling between the two sites and the rest of the chain is strong, sub-exponential scaling can be achieved. In this case, the QFI scales as $T^2$ [cf.~Eq.~\eqref{eq:qfitb2lt}]. ${\rm (b)}$ Energy spectrum of the optimal POVM. For weak coupling, the temperature-independent spectrum results in the exponential scaling of the quantum Fisher information. For strong coupling, the ground-state splits at finite temperatures resulting in gaps that scale as $T^3$. This results in the quadratic scaling of the quantum Fisher information.}
  \label{fig:weakstrong}
\end{figure*}

\subsection{Sub-exponential scaling}\label{sec:examplesB}

We consider again the tight-binding chain discussed in the last section. However, we now focus the case where one only has access to two sites of the (infinitely long) chain. In this case, all accessible observables are encoded in the reduced state that describes the two accessible sites. For simplicity, we consider the case where all POVMs that only act on the two accessible states can be implemented. As discussed in Ref.~\cite{peschel:2003}, the reduced state can be obtained from the covariance matrix. With the help of the linear spectrum approximation and in the thermodynamic limit, we find $\langle\hat{c}_j^\dagger\hat{c}_{j}\rangle=1/2$ (half filling) and
\begin{equation}
\label{eq:cov}
\begin{aligned}
\langle\hat{c}_j^\dagger\hat{c}_{j'}\rangle &=\frac{1}{\pi}\int_{-\pi/2}^{\pi/2}d\kappa \frac{\cos\left[\left(\kappa+\frac{\pi}{2}\right)(j-j')\right]}{e^{2t\kappa/k_BT}+1}\\&\simeq \frac{k_BT}{2 t}\frac{\sin\left[\frac{\pi}{2}(j-j')\right]}{\sinh\left[\pi k_BT(j-j')/(2t)\right]},
\end{aligned}
\end{equation}
where the last equality holds for $k_BT\ll t$. Following Ref.~\cite{peschel:2003}, we find the reduced state for two sites (with $j=1$ and $j=2$)
\begin{equation}
\label{eq:redstate2}
\hat{\rho}_2 = \frac{e^{-\beta\sum_{\sigma=\pm}\varepsilon_\sigma \hat{c}_\sigma^\dagger\hat{c}_\sigma}}{Z_2},
\end{equation}
with the energies
\begin{equation}
\label{eq:enred2}
\varepsilon_\pm=\pm k_BT\ln\left(\frac{t\sinh[\pi k_B T/(2t)]-k_BT}{t\sinh[\pi k_B T/(2t)]+k_BT}\right),
\end{equation}
and modes
\begin{equation}
\label{eq:statepm}
\hat{c}_\pm = \frac{1}{\sqrt{2}}\left(\hat{c}_1\pm\hat{c}_2\right).
\end{equation}
Since we consider the scenario where all POVMs that only act on the reduced state can be implemented, the relevant bound on the error in a temperature measurement is determined by the QFI of the reduced state
\begin{widetext}
	\begin{equation}
	\label{eq:qfitb2}
	\mathcal{F}_T=\frac{k_B^2}{2t^2\sinh^2[\pi k_BT/(2t)]}\frac{[\pi k_BT\cosh[\pi k_BT/(2t)]-2t\sinh[\pi k_BT/(2t)]]^2}{t^2\sinh^2[\pi k_BT/(2t)]-(k_BT)^2}.
	\end{equation}
\end{widetext}
At low temperatures, this reduces to
\begin{equation}
\label{eq:qfitb2lt}
\mathcal{F}_T=\frac{\pi^4 k_B^2}{18(\pi^2-4)}\frac{(k_BT)^2}{t^4}+\mathcal{O}[k_B^6 T^4/t^6].
\end{equation}
We thus find that the QFI of the reduced state of an infinite tight-binding chain vanishes quadratically at low temperatures.

The optimal POVM for determining temperature from the state given in Eq.~\eqref{eq:redstate2} is a measurement of the occupation numbers of the modes given in Eq.~\eqref{eq:statepm}. As discussed in Sec.~\ref{sec:sett}, we can associate an energy to each measurement outcome. This is done explicitly in App.~\ref{app:povm} and the resulting energies are plotted in Fig.~\ref{fig:weakstrong} ${\rm (b)}$. At low temperatures, we find that these energies show a ground state splitting that scales as $T^3$ which implies a Fisher information that is proportional to $T^2$ [cf.~Eq.~\eqref{eq:fisherpol}] in agreement with Eq.~\eqref{eq:qfitb2lt}. In the present example, the QFI for the full system diverges implying that temperature can in principle be measured precisely in the low $T$ limit. However, the restriction imposed on the measurements (only two sites can be addressed) necessarily implies a diverging error in low-temperature thermometry.

We can interpret the last scenario as having access to a probe that consists of two fermionic sites which are coupled to each other with a hopping term of strength $t$. This probe is then attached to a bath that consists of $N-2$ fermionic sites. The connection is made at both ends of the probe with hopping strength $t'$ [cf.~inset in Fig.~\ref{fig:weakstrong} ${\rm (a)}$]. The above analysis then corresponds to the case where $t'=t$ which corresponds to strong coupling between probe and bath (note that we made the assumption $t\gg k_BT$). It is illustrative to compare this case to the weak coupling case $t'\ll t$. Then the reduced state is given by a thermal state determined by the local Hamiltonian of the two sites
\begin{equation}
\label{eq:redstate22}
\tilde{\rho}_2 = \frac{e^{-\beta\sum_{\sigma=\pm}\tilde{\varepsilon}_\sigma \hat{c}_\sigma^\dagger\hat{c}_\sigma}}{\tilde{Z}_2},
\end{equation}
with the energies $\tilde{\varepsilon}_\pm=\mp t$, and the states given in Eq.~\eqref{eq:statepm}. The QFI of this state reads
\begin{equation}
\label{eq:qfitb22}
\mathcal{F}_T=\frac{t^2}{2k_B^2T^4}\frac{1}{\cosh^2[t/(2k_BT)]}\xrightarrow{T\rightarrow 0}\frac{2t^2}{k_B^2T^4}e^{-\frac{t}{k_B T}}.
\end{equation}
At low temperatures, it assumes the usual exponential form [cf.~Eq.~\eqref{eq:qfiltfb}] with a degeneracy factor of $g=2$. We thus find that strong coupling can outperform weak coupling at low temperatures. This is illustrated in Fig.~\ref{fig:weakstrong}. Interestingly, both for weak and for strong coupling, we find that it is the same POVM that is optimal for thermometry. As discussed in detail in App.~\ref{app:povm} and shown in Fig.~\ref{fig:weakstrong} ${\rm (b)}$, at weak coupling the energies associated to the measurement outcomes are temperature independent while at strong coupling they are temperature dependent. Furthermore, at strong coupling the energies exhibit a splitting of the ground state, allowing for the polynomial scaling of the Fisher information. {Since the optimal POVM only has four outcomes, it can be ruled out that the observed sub-exponential scaling originates from an infinite resolution as it does for the QFI of gapless systems.}

We note that a similar quadratic scaling in the strong coupling case is found in Ref.~\cite{correa:2017} for a probe consisting of a harmonic oscillator coupled to a bath of harmonic oscillators. In this case, the optimal POVM depends on the coupling strength {and has infinitely many outcomes.}

\begin{figure*}[t]
\centering
\includegraphics[width=\textwidth]{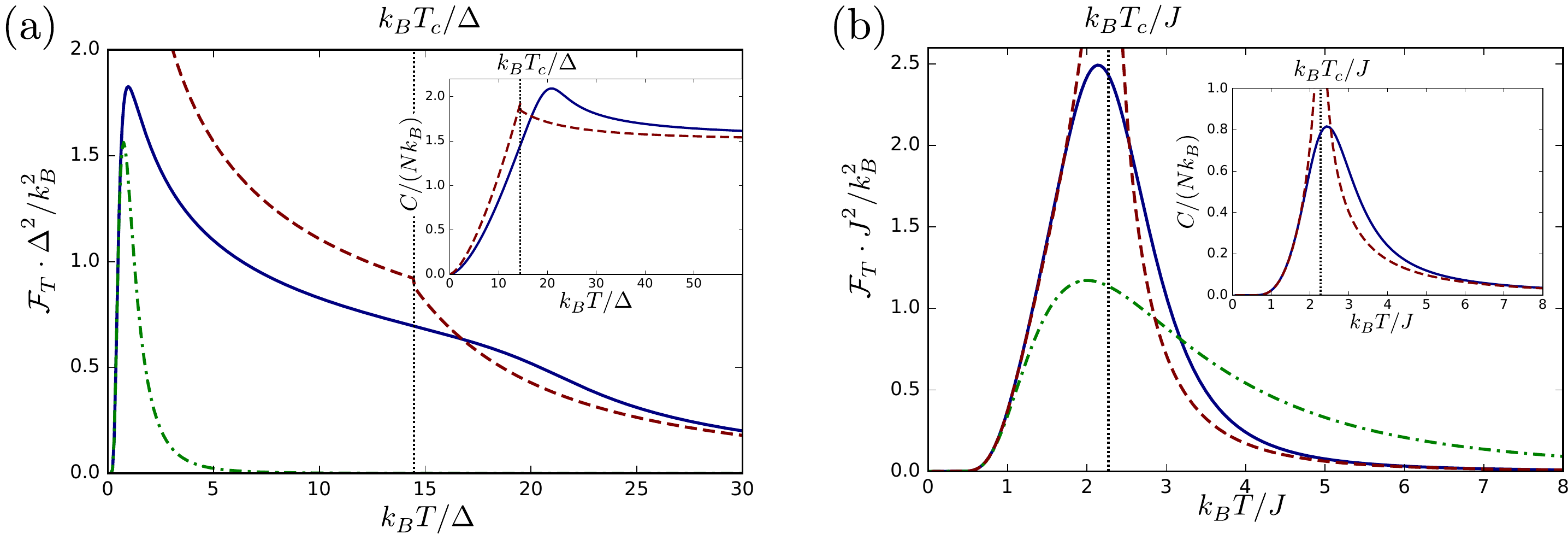}
\caption{Quantum Fisher information for systems exhibiting phase transitions. For both panels, blue (solid) shows the exact expressions for a finite system, green (dash-dotted) the corresponding low temperature limit [cf.~Eq.~\eqref{eq:qfiltfb}], and red (dashed) the thermodynamic limit $N\rightarrow\infty$. ${\rm (a)}$ Non-interacting Bose-gas with $N=100$ bosons in three dimensions. At low temperatures, the QFI follows the exponential behavior while at high temperatures it is well described by the thermodynamic limit (where $\Delta\propto 1/L\rightarrow 0$). The phase transition (Bose-Einstein condensation) gives rise to a shoulder in the QFI. The inset shows the heat capacity which has a cusp at the critical temperature $T_c$ in the thermodynamic limit [the small discontinuity is due to the approximation in Eq.~\eqref{eq:cbecht}]. ${\rm (b)}$ Square-lattice Ising model with $N=16$ spins. In this model, the QFI at low temperatures is well captured by the expression for the thermodynamic limit. This is due to the fact that the smallest gap does not vanish in the thermodynamic limit (as $J$ remains constant). In the thermodynamic limit, both the QFI as well as the heat capacity diverge at the critical temperature, allowing for precise thermometry at least in principle.}
  \label{fig:phaset}
\end{figure*}

\section{Phase transitions}
\label{sec:phaset}

Given the connection between the QFI and the heat capacity of a system [cf.~Eq.~\eqref{eq:qfic}], one could hope to make use of phase transitions, where the heat capacity can show a singular behavior, in order to enhance the precision of thermometry. Indeed the connection between critical behavior and metrology was investigated before \cite{zanardi:2008}. Here we investigate two phase transitions. Bose-Einstein condensation (BEC), where the heat capacity exhibits a maximum at the critical temperature, and the Ising model, where the heat capacity diverges at the phase transition to a magnetically ordered state. We note that while phase transitions can enhance thermometry at the critical temperature, there seems to be no direct influence on the scaling of the Fisher information as $T\rightarrow0$. 


\subsection{Bose-Einstein condensation}

To investigate BEC, we consider a non-interacting Bose gas in three dimensions with a fixed density. The chemical potential then becomes temperature-dependent and is determined by
\begin{equation}
\label{eq:chemt}
N=\sum_k\frac{1}{e^{\beta[\varepsilon_k-\mu(T)]}\pm 1},
\end{equation}
where $N$ denotes the number of particles. In the thermodynamic limit, where $N, L\rightarrow\infty$ with the density $N/L^3$ fixed, the Bose gas condenses into the ground state at the critical temperature
\begin{equation}
\label{eq:bectc}
k_BT_c=\frac{2\pi}{m}\left(\frac{N}{L^3\zeta(3/2)}\right)^{2/3}.
\end{equation}
The heat capacity exhibits a cusp at the critical temperature and reads
\begin{equation}
\label{eq:cbeclt}
\frac{C}{k_BN}=\frac{15}{4}\frac{\zeta(5/2)}{\zeta(3/2)}\left(\frac{T}{T_c}\right)^{3/2},\hspace{.25cm}{\rm for}\,\,T\leq T_c,
\end{equation}
and
\begin{equation}
\label{eq:cbecht}
\frac{C}{k_BN}\simeq\frac{3}{2}\left[1+\frac{\zeta(3/2)}{2^{7/2}}\left(\frac{T_c}{T}\right)^{3/2}\right],\hspace{.25cm}{\rm for}\,\,T> T_c.
\end{equation}
Note that the last expression is only approximative (see Ref.~\cite{london:book} for a detailed discussion).

Figure \ref{fig:phaset} ${\rm (a)}$ shows the QFI of a finite system together with the thermodynamic limit obtained from Eqs.~\eqref{eq:cbeclt} and \eqref{eq:cbecht}. While the maximum in the heat capacity leads to a shoulder in the QFI, the system is much more sensitive to temperatures which are of the order of the smallest gap. For the finite system, we numerically evaluated Eq.~\eqref{eq:qfib} with the chemical potential determined by Eq.~\eqref{eq:chemt} with $N=100$.

\subsection{Ising model}
Here we consider a square-lattice Ising model in two dimensions described by the Hamiltonian
\begin{equation}
\label{eq:ising}
\hat{H}=-J\sum_{\langle i,j\rangle}\hat{\sigma}_z^i\hat{\sigma}_z^j,
\end{equation}
where the sum goes over nearest neighbors. This model exhibits a phase transition (in the thermodynamic limit) to a magnetically ordered state [(anti)ferro-magnetic for $J>0$ ($J<0$)] at the critical temperature
\begin{equation}
k_BT_c = \frac{2J}{\ln(1+\sqrt{2})}.
\end{equation}
In the thermodynamic limit, the heat capacity reads \cite{onsager:1944}
\begin{widetext}
	\begin{equation}
	\label{eq:cising}
	\begin{aligned}
	\frac{C}{k_BN}=\frac{4}{\pi}\left[\frac{J}{k_BT}\coth(2J/k_BT)\right]^2\left\{K_1(z)-E_1(z)-\left[1-\tanh^2(2J/k_BT)\right]\left[\frac{\pi}{2}+[2\tanh^2(2J/k_BT)-1]K_1(z)\right]\right\},
	\end{aligned}
	\end{equation}
\end{widetext}
where $N$ denotes the number of spins and we introduced
\begin{equation}
\label{eq:isingz}
z=2\frac{\sinh(2J/k_BT)}{\cosh^2(2J/k_BT)},
\end{equation}
and the complete elliptic integrals of the first and second kind
\begin{equation}
\label{eq:compell}
\begin{aligned}
&K_1(z)=\int_{0}^{\pi/2}d\phi\frac{1}{\sqrt{1-z^2\sin^2(\phi)}},
\\&E_1(z)=\int_{0}^{\pi/2}d\phi\sqrt{1-z^2\sin^2(\phi)}.
\end{aligned}
\end{equation}

In Fig.~\ref{fig:phaset} ${\rm (b)}$ we plot the QFI for a finite Ising lattice together with the thermodynamic limit obtained from Eq.~\ref{eq:cising}. For the finite system, we considered a square lattice of $N=16$ spins with periodic boundary conditions in both directions (i.e., a lattice arranged on a torus). We numerically evaluated the variance of the Hamiltonian in a thermal state to determine the QFI [cf.~Eq.~\eqref{eq:qfi}].
We note that already for such a small system, the QFI is substantially enhanced at $T_c$ as compared to the low-temperature approximation. In this system, the effect of the gapped structure (the gap is equal to $8J$, corresponding to flipping a single spin) and the phase transition add up to a large peak in the QFI.

\section{Conclusions}
\label{sec:conclusions}

We investigated fundamental limitations on low-temperature quantum thermometry {with finite resolution}. We developed an approach where restrictions from both the sample and the measurement are taken into account. Formally, we associate an energy as well as a probability to each measurement outcome, which fully determine the Fisher information of the corresponding POVM. This leads to a classification of POVMs, corresponding to different low-temperature scalings of the measurement error. While most POVMs lead to an exponentially diverging error, this is not the case in general. The temperature dependence of the associated energies can result in a sub-exponential scaling of the error, whenever the ground state exhibits a degeneracy at zero temperature which is lifted at finite temperatures. {Extending our results to POVMs with associated spectra that are approximatively gapless, such that the finite-resolution assumption breaks down, provides an interesting avenue of future research.} {We note that our bounds are no longer valid if the POVM elements are allowed to become temperature-dependent as in Refs.~\cite{paris:2016,mukherjee:2018}.}

{As illustrated in Fig.~\ref{fig:1}, the introduced spectra provide a visual tool to distinguish between scenarios where the measurement error scales exponentially and scenarios where a polynomial scaling can be obtained. Our approach allowed us to show that infinite resolution is \textit{not} required for observing a sub-exponential scaling in the measurement error, and provides an intuitive understanding of the origin of this advantage. In particular, for the specific system of a fermionic tight-binding chain with access to only two sites, we find an error that scales as $\delta T^2\propto 1/T^2$. An identical scaling was found in Ref.~\cite{correa:2017}, considering a bosonic probe strongly coupled to an infinite bath of harmonic oscillators, and a measurement with infinite resolution. It would be interesting to investigate the results of Ref.~\cite{correa:2017} in the light of our approach, and see whether the requirement of infinite resolution can be relaxed.}

{Furthermore, we derived a fundamental bound on the precision of quantum thermometry assuming only that measurements have finite resolution. The same bound can be obtained from a different assumption based on the third law of thermodynamics. Interestingly, this bound is not saturated by any of the known examples of realistic quantum thermometry, and in principle allows for better schemes. A central question is now to determine the best possible scaling which is physically achievable. This scaling must lie somewhere in between the lower bound and the quadratic scaling provided by our explicit example. In the case a scaling better than quadratic is possible, we believe that our formalism represents a promising approach to identify explicit schemes. Intriguingly, our lower bound in principle allows the absolute error to vanish as $T \rightarrow 0$, and it would be interesting to see if this is actually possible in a physical system. {We finally note that throughout this paper we quantified thermometry through the obtained information. A complementary measure is provided by the disturbance of the procedure \cite{seveso:2018}. The generalization of our approach to include disturbance is left open for future work.}}

\paragraph{Note added.} After completion of this manuscript, we became aware of related work \cite{hovhannisyan:2018}, which considers thermometry on gapless systems by studying coupled harmonic oscillators.

\paragraph{Acknowledgements.} We thank Paul Skrzypczyk and Mart\'i Perarnau-Llobet for discussions. We acknowledge support from the Swiss National Science Foundation (Grant No. $200021 \_ 169002$, and QSIT).

\bibliographystyle{quantum_ph}
\bibliography{biblio}

\onecolumn\newpage
\appendix

\section{Chemical potential}
\label{app:mu}
In this section, we consider systems described by the grand canonical ensemble, i.e, we consider states of the form
\begin{equation}
\label{eq:grandcan}
\hat{\rho}_\mu=\frac{e^{-\beta(\hat{H}-\mu\hat{N})}}{Z_\mu},
\end{equation}
where $[\hat{H},\hat{N}]=0$ and the partition sum reads $Z_\mu={\rm Tr}\{\exp{[-\beta(\hat{H}-\mu\hat{N})]}\}$. Here we only consider a single chemical potential. Extensions to multiple particle species, each with their own chemical potential, are straightforward. For a temperature-independent chemical potential, we can simply replace $\hat{H}\rightarrow\hat{H}-\mu\hat{N}$ in Sec.~\ref{sec:sett}. The bounds derived in Sec.~\ref{sec:bounds} are then still valid but the energies $E_m$ are defined as
\begin{equation}
\label{eq:emu}
E_m=\frac{1}{p_m}{\rm Tr}\left\{\hat{\Pi}_m(\hat{H}-\mu\hat{N})\hat{\rho}_T\right\}.
\end{equation}
For the QFI, we find in this case
\begin{equation}
\label{eq:qfimu}
\mathcal{F}_T=\frac{\langle (\hat{H}-\mu\hat{N})^2\rangle -\langle (\hat{H}-\mu\hat{N})\rangle}{k_B^2T^4}=\frac{\partial_T\langle (\hat{H}-\mu\hat{N})\rangle}{k_B T^2}=\frac{\partial_\beta^2\ln Z_\mu}{k_B^2T^4}= \frac{\partial_T S}{k_BT} =\frac{C}{k_BT^2}.
\end{equation}
Therefore, the bound in Eq.~\eqref{eq:bound3rdlaw} still holds. We note that in this case, the optimal POVM is given by a projective measurement of energy and particle number which leads to temperature-independent $E_m$ \cite{marzolino:2013}.

For a temperature-dependent chemical potential, using the definitions in Eq.~\eqref{eq:pm} and \eqref{eq:fisher}, the Fisher information can still be written as the variance of the energies associated to the POVM elements [cf.~Eq.~\eqref{eq:fishercl}]. However, these energies now read
\begin{equation}
\label{eq:emut}
E_m=\frac{1}{p_m}{\rm Tr}\left\{\hat{\Pi}_m[\hat{H}-(\mu-T\partial_T\mu)\hat{N}]\hat{\rho}_T\right\}.
\end{equation}
Note that they can be temperature dependent even for the optimal POVM which is still given by a projective measurement of both energy and particle number. This implies that even the QFI for a gapped system can exhibit a sub-exponential scaling at low temperatures. For this to happen, the lowest energy $E_0$ needs to be degenerate at $T=0$ and this degeneracy must be lifted at finite temperatures. For the optimal POVM, the energies are of the form 
\begin{equation}
\label{eq:enoptmu}
E_m=\mathcal{E}_m-(\mu-T\partial_T\mu)\mathcal{N}_m,
\end{equation}
where $\mathcal{E}_m$ and $\mathcal{N}_m$ are the eigenvalues of $\hat{H}$ and $\hat{N}$. This implies that a sub-exponential scaling can only be achieved if the states involved in the ground-state degeneracy at $T=0$ correspond to different particle numbers. We therefore find that a temperature-dependent chemical potential together with a particle-number uncertainty in the ground-state manifold are necessary ingredients for a QFI that scales sub-exponentially. We note that if the chemical potential can be Taylor-expanded around $T=0$, then the ground-state degeneracy is lifted at least proportionally to $T^2$. Thus Eq.~\eqref{eq:fisherpol} applies and the QFI scales at least with $T^0$.

A temperature dependent chemical potential also alters Eq.~\eqref{eq:qfimu} and we find
\begin{equation}
\label{eq:qfimut}
\mathcal{F}_T=\frac{\partial_\beta^2\ln Z_\mu}{k_B^2T^4}-\frac{(\partial_T\mu)^2}{k_BT}\langle \hat{N}\rangle=\frac{C}{k_BT^2}+(\partial_T\mu)(\partial_T\langle \hat{N}\rangle).
\end{equation}
The QFI is thus directly proportional to the heat capacity if either the chemical potential or the mean particle number is fixed as a function of temperature.

We close this section with a brief discussion on the scenario where both temperature as well as chemical potential are fixed but unknown parameters that are to be determined by a measurement. This corresponds to the problem of multi-parameter estimation where the variances of any measurement are bounded by
\begin{equation}
\label{eq:varmp}
\begin{pmatrix}
\delta T^2 & \delta T\delta\mu\\
\delta T\delta\mu & \delta\mu^2
\end{pmatrix}\geq \frac{1}{\nu}\mathcal{F}^{-1},
\end{equation}
with the QFI matrix given by \cite{marzolino:2013}
\begin{equation}
\label{eq:fishermat}
\mathcal{F}=\begin{pmatrix}
\mathcal{F}_{T} & \mathcal{F}_{T\mu}\\
\mathcal{F}_{T\mu} & \mathcal{F}_\mu
\end{pmatrix},
\end{equation}
where
\begin{equation}
\label{eq:fishermatexpl}
\begin{aligned}
\mathcal{F}_T&=\frac{\partial_\beta^2\ln Z_\mu}{k_B^2 T^4}=\frac{\partial_T S}{k_BT}=\frac{\partial_T\langle \hat{H}-\mu\hat{N}\rangle}{k_BT^2},\\
\mathcal{F}_\mu &= \partial^2_\mu \ln Z_\mu = \frac{\partial_\mu\langle \hat{N}\rangle }{k_B T},\\
\mathcal{F}_{T\mu} &= \frac{1}{k_BT}\partial_\beta \frac{1}{\beta}\partial_\mu\ln Z_\mu = \frac{\partial_T\langle \hat{N}\rangle}{k_B T}.
\end{aligned}
\end{equation}
We note that if we have perfect knowledge of $\mu$, then we can set $\delta\mu=0$ and we recover the results for a temperature-independent chemical potential given in Eq.~\eqref{eq:qfimu}.
In the next section, we reproduce the above relations by a classical calculation.

\section{Classical bound on thermometry}
\label{app:classical}
Here we reproduce the calculation on fluctuations of thermodynamic quantities given in Ref.~\cite{landaulifshitz}. Interestingly, the variances found for these fluctuations are exactly the minimal variances given in Eqs.~\eqref{eq:varmp}-\eqref{eq:fishermatexpl}. Intuitively, the variance of any temperature measurement is bounded from below by the variance in the intrinsic temperature fluctuations.

As in Ref.~\cite{landaulifshitz}, we consider a large, closed system with a well defined energy $E_t$. The probability of finding the total system in a given configuration is then determined by the entropy as
\begin{equation}
\label{eq:probatot}
w\propto\delta(E-E_t)e^{S_t/k_B}=\delta(E-E_t)e^{(S_0+\Delta S_t)/k_B},
\end{equation}
where $S_t$ denotes the total entropy and $S_0$ its maximal value (i.e. $\Delta S_t\leq 0$). We now consider a small part of the total system which is still macroscopic in size. We denote this part as the system whereas all the rest of the total system is denoted as the bath. We make the assumption that the system locally equilibrates on time-scales that are much faster than all other time-scales of the problem. In this case, we can restrict the analysis to local equilibrium states which are described by a temperature $T$, a chemical potential $\mu$, and an entropy $S$. All these quantities depend on the energy $E$ and the particle number $N$ in the system. Due to energy and particle exchange with the bath, the thermodynamic quantities that describe the system will fluctuate. In particular, the energy deviation from its equilibrium value reads
\begin{equation}
\label{eq:enfluc}
\Delta E = -T_B\Delta S_B-\mu_B\Delta N_B=-T_B\Delta S_t+T_B\Delta S+\mu_B\Delta N,
\end{equation}
where $\Delta S_B$ and $\Delta N_B$ are the entropy and particle number deviation in the bath. For the last equality, we used $S_t=S_B+S$ and $\Delta N_B=-\Delta N$ and all quantities without subscript denote the system. From Eq.~\eqref{eq:enfluc}, we find
\begin{equation}
\label{eq:delst}
\Delta S_t=-\frac{1}{T_B}(\Delta E-T_B\Delta S-\mu_B\Delta N)
\simeq -\frac{1}{2T}(\Delta S\Delta T+\Delta N\Delta\mu).
\end{equation}
To obtain the second equality, we used $T_B=T+\Delta T$ and $\mu_B=\mu+\Delta\mu$ with
\begin{equation}
\label{eq:delt}
T=\left(\frac{\partial E}{\partial S}\right)_N,\hspace{1cm}\mu=\left(\frac{\partial E}{\partial N}\right)_S,
\end{equation}
and neglected all higher order terms in $\Delta X$ by expanding $\Delta E$ ($\Delta T$, $\Delta \mu$) up to second (first) order in $\Delta S$ and $\Delta N$ \cite{landaulifshitz}.

Taking $T$ and $\mu$ as independent variables and using the Maxwell relation
\begin{equation}
\label{eq:maxwell}
\partial_\mu S=\partial_T N,
\end{equation}
where we keep $\mu$ constant when taking the derivative with respect to $T$  and vice versa, we find
\begin{equation}
\label{eq:delsfin}
\Delta S_t = -\frac{1}{2T}\begin{pmatrix}
\Delta T  &\Delta \mu
\end{pmatrix}\begin{pmatrix}
\partial_T S &  \partial_T N\\
\partial_T N & \partial_\mu N
\end{pmatrix}
\begin{pmatrix}
\Delta T\\
\Delta\mu
\end{pmatrix}.
\end{equation}
Identifying $N$ with $\langle \hat{N}\rangle$, we find that the fluctuations of the total entropy is governed by the QFI given in Eq.~\eqref{eq:fishermatexpl}. The probability of observing a temperature deviation $\Delta T$ and a chemical potential deviation $\Delta \mu$ from their equilibrium values is thus given by [cf.~Eq.~\eqref{eq:probatot}]
\begin{equation}
w(\Delta T,\Delta \mu)\propto\exp\left[-\frac{1}{2}\begin{pmatrix}
\Delta T  &\Delta \mu
\end{pmatrix}\mathcal{F}\begin{pmatrix}
\Delta T\\
\Delta\mu
\end{pmatrix}\right].
\end{equation}
Thus temperature and chemical potential fluctuate with variances given by the inverse of the QFI matrix. Any measurement with the aim of determining $T$ and $\mu$ must thus result in variances that are strictly bigger as expressed by the Cram\'er-Rao bound in Eq.~\eqref{eq:varmp}.

\section{Thermodynamic limit for massive, non-interacting particles}
\label{app:tdlimit}
In this section, we consider massive, non-interacting particles enclosed in a cubic container with volume $L^d$, where $d$ denotes the spatial dimension. The single-particle energies are given by $\varepsilon_k=k^2/(2m)$, where the quantum number $k$ can take on the values 
\begin{equation}
\label{eq:kapp}
k=\frac{\pi}{L}\sqrt{\sum_{i=1}^d n_i^2},
\end{equation}
with $n_i$ being positive integers. Here we are interested in the thermodynamic limit, i.e., the limit $L\rightarrow \infty$ (keeping the density finite), where $k$ becomes a continuous variable. The QFI for such a system is given in Eqs.~\eqref{eq:qfib} and \eqref{eq:qfif} for bosons and fermions respectively. In the thermodynamic limit, we can replace the sum over $k$ with an integral resulting in
	\begin{equation}
	\label{eq:fisherbcontfd}
	\mathcal{F}_T^B =\frac{\nu_d}{T^{2-d/2}}\left(\frac{\sqrt{mk_B}L}{\pi}\right)^{d}\int_{-\mu/(2k_BT)}^{\infty}dx \left[x+\partial_T\mu(T)/(2k_B)\right]^2\frac{\left(x+\mu/(2k_BT)\right)^{d/2-1}}{\sinh^2(x)},
	\end{equation}
	and
	\begin{equation}
	\label{eq:fisherfcontfd}
	\mathcal{F}_T^F =\frac{\nu_d}{T^{2-d/2}}\left(\frac{\sqrt{mk_B}L}{\pi}\right)^{d}\int_{-\mu/(2k_BT)}^{\infty}dx \left[x+\partial_T\mu(T)/(2k_B)\right]^2\frac{\left(x+\mu/(2k_BT)\right)^{d/2-1}}{\cosh^2(x)},
	\end{equation}
	where $\nu_1=1$, $\nu_2=\pi$, and $\nu_3=2\pi$. As expected, the QFI scales with the volume $L^d$ \cite{marzolino:2013}. In two dimensions, the above integrals can be solved analytically yielding
	\begin{equation}
	\label{eq:fisherbcontd2fd}
	\begin{aligned}
	\mathcal{F}_T^B = &\frac{m k_B}{\pi T}L^2\left[{\rm Li}_2(e^{\beta\mu})+\frac{\mu}{k_BT}\ln\left(1-e^{\beta\mu}\right)-\frac{(\mu-T\partial_T\mu)^2}{4k_B^2T^2}\left[1+\coth\left(\mu/(2k_BT)\right)\right]\right]\\&-\frac{\mu\partial_T\mu}{2k_B^2T^4}-\frac{\partial_T\mu}{k_B}\ln\left[2\sinh(-\mu/(2k_BT))\right],
	\end{aligned}
	\end{equation}
	and
	\begin{equation}
	\label{eq:fisherfcontd2fd}
	\begin{aligned}
	\mathcal{F}_T^F =& \frac{m k_B}{\pi T}L^2\left[-{\rm Li}_2(-e^{\beta\mu})-\frac{\mu}{k_BT}\ln\left(1+e^{\beta\mu}\right)+\frac{(\mu-T\partial_T\mu)^2}{4k_B^2T^2}\left[1+\tanh\left(\mu/(2k_BT)\right)\right]\right]\\&+\frac{\mu\partial_T\mu}{2k_B^2T^4}+\frac{\partial_T\mu}{k_B}\left[\ln\left[\cosh(\mu/(2k_BT))\right]+\ln(2)\right],
	\end{aligned}
	\end{equation}
where ${\rm Li}_2(x)$ denotes the polylogarithm of order two. Equation \eqref{eq:fisherfcontd2fd} is plotted in Fig.~\ref{fig:examples_gapped} ${\rm (b)}$.

For a temperature-independent chemical potential $\mu$, we can find the low temperature behavior of Eqs.~\eqref{eq:fisherbcontd2fd} and \eqref{eq:fisherfcontd2fd} in all dimensions. To this end, we distinguish three regimes.

\subsection{$\mu=0$}
In this case, we use
\begin{equation}
\label{eq:intsinh}
\int_{0}^{\infty}dx\frac{x^{\frac{d}{2}+1}}{\sinh^2(x)}=\frac{1}{2^{\frac{d}{2}}}\Gamma\left(\frac{d}{2}+2\right)\zeta\left(\frac{d}{2}+1\right),
\end{equation}
and
\begin{equation}
\label{eq:intcosh}
\int_{0}^{\infty}dx\frac{x^{\frac{d}{2}+1}}{\cosh^2(x)}=\frac{2^{\frac{d}{2}}-1}{2^d}\Gamma\left(\frac{d}{2}+2\right)\zeta\left(\frac{d}{2}+1\right),
\end{equation}
where $\Gamma(x)$ denotes the gamma function and $\zeta(x)$ the Riemann zeta function. With the help of the above identities, we find
\begin{equation}
\label{eq:qfimu0b}
\mathcal{F}_T^B = \frac{\nu_d}{2^{\frac{d}{2}}T^{2-d/2}}\left(\frac{\sqrt{mk_B}L}{\pi}\right)^{d}\Gamma\left(\frac{d}{2}+2\right)\zeta\left(\frac{d}{2}+1\right),
\end{equation}
and
\begin{equation}
\label{eq:qfimu0f}
\mathcal{F}_T^F = \frac{(2^{\frac{d}{2}}-1)\nu_d}{2^dT^{2-d/2}}\left(\frac{\sqrt{mk_B}L}{\pi}\right)^{d}\Gamma\left(\frac{d}{2}+2\right)\zeta\left(\frac{d}{2}+1\right).
\end{equation}
We find that the QFI is proportional to $T^{\frac{d}{2}-2}$ in contrast to photons [cf.~Eq.~\eqref{eq:qficontph}], where we find the scaling $T^{d-2}$. We note that these expressions for the QFI are valid at all temperatures (but they are restricted to $\mu=0$). We further note that for a Bose gas with fixed density, the chemical potential is pinned to $\mu=0$ below the critical temperature. Indeed, multiplying Eq.~\eqref{eq:qfimu0b} with $T^2/N$ and setting $d=3$, we recover Eq.~\eqref{eq:cbeclt}.

\subsection{$\mu<0$}
For negative chemical potentials, we perform a shift $x\rightarrow x-\mu/(2k_BT)$ in the integrals of Eqs.~\eqref{eq:fisherbcontfd} and \eqref{eq:fisherfcontfd}. We then make the approximation
\begin{equation}
\label{eq:approxsinhcosh}
\sinh[x+|\mu|/(2k_BT)]\simeq \cosh[x+|\mu|/(2k_BT)]\simeq \frac{1}{2}e^{x}e^{|\mu|/(2k_BT)},
\end{equation}
which is valid in the limit $k_BT\ll |\mu|$. We further use the identity
\begin{equation}
\label{eq:identgamma}
\Gamma\left(\frac{d}{2}\right)=2^\frac{d}{2}\int_{0}^{\infty}dx x^{\frac{d}{2}-1}e^{-2x},
\end{equation}
resulting in the QFI
\begin{equation}
\label{eq:qfimuneg}
\mathcal{F}_T=\nu_d\frac{\Gamma\left(\frac{d}{2}\right)}{2^{\frac{d}{2}}}\left(\frac{\sqrt{mk_B}L}{\pi}\right)^{d}\frac{\mu^2e^{-|\mu|/(k_BT)}}{k_B^2T^{4-\frac{d}{2}}},
\end{equation}
which holds both for bosons and for fermions in the limit $k_BT\ll|\mu|$. Due to the negative chemical potential, there is a gap of magnitude $|\mu|$ between the ground-state (no particles), and the first excited state (presence of a single particle). This results in an exponential scaling of the QFI similar to what we found for a discrete spectrum [cf.~Eq.~\eqref{eq:qfiltfb}]. The fact that there is a continuum of states above the gap, modifies the scaling in the denominator [$T^{4}$ in Eq.~\eqref{eq:qfiltfb} vs. $T^{4-\frac{d}{2}}$ in Eq.~\eqref{eq:qfimuneg}] similar to what we found if the excited states exhibit a temperature dependence [cf.~Eq.~\eqref{eq:fisherexp2}].

\subsection{$\mu>0$}
This regime is exclusively of relevance for fermions since $\mu\leq 0$ for bosons (in order to have finite occupation numbers). In the limit $k_BT\ll\mu$, we assume that the integrand in Eq.~\eqref{eq:fisherfcontfd} is only non-zero for $x\ll\mu/k_BT$. We can then extend the integral to go from minus infinity to infinity and approximate $x+\mu/(2k_BT)\simeq\mu/(2k_BT)$. With the identity
\begin{equation}
\label{eq:idcoshxs}
\int_{-\infty}^{\infty}dx\frac{x^2}{\cosh^2(x)}=\frac{\pi^2}{6},
\end{equation}
we then find
\begin{equation}
\label{eq:qfifmup}
\mathcal{F}_T^F=\frac{\nu_d}{2^\frac{d}{2}}\frac{\pi^2}{3}\left(\sqrt{\frac{m}{\mu}}\frac{L}{\pi}\right)^d\frac{k_B}{\mu T}.
\end{equation}
We thus find a QFI that diverges as $1/T$ for all dimensions as long as the chemical potential is within the energy spectrum. This scaling implies a heat capacity that vanishes linearly in temperature as expected for free fermions \cite{ashcroft:book}.

\section{Thermometry with two fermionic sites: optimal POVM}
\label{app:povm}
In this section, we consider the optimal POVM for thermometry given the states in Eqs.~\eqref{eq:redstate2} and \eqref{eq:redstate22}. The first state is obtained by tracing out all but two sites of a tight-binding chain described by the Hamiltonian in Eq.~\eqref{eq:hamtb}. The second state is obtained by weakly coupling two fermionic sites (coupled to each other with tunnel coupling $t$) to a thermal bath. In both cases we focus on half-filling, i.e. $\mu=\varepsilon$. Both states are diagonal in the basis given in Eq.~\eqref{eq:statepm}. The optimal POVM is thus the same for both states and is given by measuring the occupation numbers in the diagonal basis. The corresponding POVM elements read
\begin{equation}
\label{eq:povmtb2}
\begin{aligned}
&\hat{\Pi}_0=\left(1-\hat{c}_+^\dagger\hat{c}_+\right)\left(1-\hat{c}_-^\dagger\hat{c}_-\right),\\&\hat{\Pi}_\pm=\hat{c}_\pm^\dagger\hat{c}_\pm\left(1-\hat{c}_\mp^\dagger\hat{c}_\mp\right),\\&\hat{\Pi}_2=\hat{c}_+^\dagger\hat{c}_+\hat{c}_-^\dagger\hat{c}_-.
\end{aligned}
\end{equation}
As discussed in Sec.~\ref{sec:sett}, we can associate probabilities [cf.~Eq.~\eqref{eq:pm}] and energies [cf.~Eq.~\eqref{eq:em}] to the POVM elements. For the reduced state of the tight-binding chain, we find
\begin{equation}
\label{eq:probatb2}
\begin{aligned}
&p_0=p_2=\frac{1}{4}-\frac{(k_BT)^2}{4t^2}\frac{1}{\sinh^2[\pi k_BT/(2t)]},\\&
p_\pm=\left[\frac{1}{2}\pm\frac{k_BT}{2t}\frac{1}{\sinh[\pi k_BT/(2t)]}\right]^2,
\end{aligned}
\end{equation}
and
	\begin{equation}
	\label{eq:emtb2}
	\begin{aligned}
	&{E}_0={E}_2=\frac{(k_BT)^3}{t\sinh[\pi k_BT/(2t)]}\frac{\pi k_BT\cosh[\pi k_BT/(2t)]-2t\sinh[\pi k_BT/(2t)]}{t^2\sinh^2[\pi k_BT/(2t)]-(k_BT)^2},
	\\&{E}_\pm=\mp\frac{(k_BT)^2}{t\sinh[\pi k_BT/(2t)]}\frac{\pi k_BT\cosh[\pi k_BT/(2t)]-2t\sinh[\pi k_BT/(2t)]}{t\sinh[\pi k_BT/(2t)]\pm k_BT}.
	\end{aligned}
	\end{equation}
With the help of Eq.~\eqref{eq:fishercl}, we recover the QFI in Eq.~\eqref{eq:qfitb2} which confirms that the POVM in Eq.~\eqref{eq:povmtb2} is indeed optimal for determining temperature if one only has access to the reduced state.
At low temperatures, the probabilities and energies reduce to
\begin{equation}
\label{eq:probatb2lt}
\begin{aligned}
&p_0=p_2=\frac{1}{4}-\frac{1}{\pi^2}+\mathcal{O}[(k_BT/t)^2],\\&p_\pm = \left(\frac{1}{2}\pm\frac{1}{\pi}\right)^2+\mathcal{O}[(k_BT/t)^2],
\end{aligned}
\end{equation}
and
\begin{equation}
\label{eq:emtb2lt}
\begin{aligned}
&
{E}_0={E}_2=\frac{2\pi^2}{3(\pi^2-4)}\frac{(k_BT)^3}{t^2}+\mathcal{O}[(k_BT)^5/t^4],\\&{E}_\pm=\mp\frac{\pi^2}{3(\pi\pm 2)}\frac{(k_BT)^3}{t^2}+\mathcal{O}[(k_BT)^5/t^4].
\end{aligned}
\end{equation}
We thus find that the energies are degenerate at $T=0$. At finite temperatures, gaps open up that scale as $T^3$, giving rise to the polynomial low temperature behavior of the quantum Fisher information discussed in the main text and illustrated in Fig.~\ref{fig:weakstrong}.

For the weakly coupled system, we find the probabilities
\begin{equation}
\label{eq:probatb22}
\begin{aligned}
&p_0=p_2=(2+2\cosh[t/(k_BT)])^{-1},\\&
p_\pm=\left[e^{\mp t/(k_BT)}+1\right]^{-2},
\end{aligned}
\end{equation}
and the energies
\begin{equation}
\label{eq:emtb22}
{E}_0={E}_2=0,
\hspace{.5cm}{E}_\pm=\varepsilon_\pm=\mp t.
\end{equation}
With the help of Eq.~\eqref{eq:fishercl}, we recover the QFI in Eq.~\eqref{eq:qfitb22}. Since the above energies are temperature independent, we find an exponential scaling of the Fisher information at low temperatures in accordance with Eq.~\eqref{eq:fisherexp}.

\end{document}